\documentclass[
amsmath, %
floatfix, %
twocolumn, %
showpacs, %
reprint, %
superscriptaddress, %
prb, %
aps, %
]{revtex4-1}
\usepackage[T1]{fontenc}
\usepackage[utf8]{inputenc}
\usepackage{graphicx}
\usepackage{latexsym}
\usepackage{amsthm}
\usepackage{color,xcolor}
\usepackage{mathtools}

\usepackage[tighter]{newtxtext}
\usepackage[subscriptcorrection,nosymbolsc,smallerops,bigdelims]{newtxmath}
\DeclareMathAlphabet{\mathcal}{OMS}{cmsy}{m}{n}

\DeclareMathAlphabet{\mathcalb}{OMS}{cmsy}{b}{n}

\usepackage{bm}

\usepackage{floatrow}
\usepackage{siunitx}
\usepackage{hyperref}
\hypersetup{
	colorlinks,
	linkcolor={blue!100!black},
	citecolor={blue!100!black},
	urlcolor={blue!100!black}
}
\usepackage{xfrac}
\usepackage{nicefrac}

\newcommand{\ket}[1]{\left| #1 \right>}

\newcommand{\matrixel}[3]{\left< #1 \vphantom{#2#3} \right|#2 \left| #3 \vphantom{#1#2} \right>}
 
\newcommand{\lr}[1]{\!\left(#1\right)}

\newlength{\Rwidth}
\newlength{\Rheight}
\newcommand{\kp}{\bm{k}{\cdot}\bm{p}}

\mathchardef\mhyphen="2D
\newcommand{\ignore}[1]{}
\newcommand{\sla}{\!\!/\!\!}

\newcommand{\smket}[1]{\lvert #1 \rangle} 
 
\newcommand{\mfrac}[2]{#1/#2}

\makeatletter
\newcommand{\raisemath}[1]{\mathpalette{\raisem@th{#1}}}
\newcommand{\raisem@th}[3]{\raisebox{#1}{$#2#3$}}
\makeatother

\newcommand{\mr}[1]{\mathrm{#1}}
\newcommand{\ii}{\mathrm{i}}

\newcommand{\upemhf}{\vspace{-0.65em}}

\begin{document}
	\title{Exciton lifetime and emission polarization dispersion in strongly in-plane asymmetric nanostructures}
	
	\author{M.~Gawe{\l}czyk}
		\email{michal.gawelczyk@pwr.edu.pl}
		\affiliation{Laboratory for Optical Spectroscopy of Nanostructures, Department of Experimental Physics, Faculty of Fundamental Problems of Technology, Wroc\l{}aw University of Science and Technology, 50-370 Wroc\l{}aw, Poland}
		\affiliation{Department of Theoretical Physics, Faculty of Fundamental Problems of Technology, Wroc\l{}aw University of Science and Technology, 50-370 Wroc\l{}aw, Poland}
		
	\author{M.~Syperek}
		\email{marcin.syperek@pwr.edu.pl}
		\affiliation{Laboratory for Optical Spectroscopy of Nanostructures, Department of Experimental Physics, Faculty of Fundamental Problems of Technology, Wroc\l{}aw University of Science and Technology, 50-370 Wroc\l{}aw, Poland}
		
	\author{A.~Mary\'nski}
		\affiliation{Laboratory for Optical Spectroscopy of Nanostructures, Department of Experimental Physics, Faculty of Fundamental Problems of Technology, Wroc\l{}aw University of Science and Technology, 50-370 Wroc\l{}aw, Poland}
		
 	\author{P.~Mrowi\'nski}
		\affiliation{Laboratory for Optical Spectroscopy of Nanostructures, Department of Experimental Physics, Faculty of Fundamental Problems of Technology, Wroc\l{}aw University of Science and Technology, 50-370 Wroc\l{}aw, Poland}

	\author{{\L}.~Dusanowski}
		\affiliation{Laboratory for Optical Spectroscopy of Nanostructures, Department of Experimental Physics, Faculty of Fundamental Problems of Technology, Wroc\l{}aw University of Science and Technology, 50-370 Wroc\l{}aw, Poland}

	\author{K.~Gawarecki}
		\affiliation{Department of Theoretical Physics, Faculty of Fundamental Problems of Technology, Wroc\l{}aw University of Science and Technology, 50-370 Wroc\l{}aw, Poland}
		
	\author{J.~Misiewicz}
		\affiliation{Laboratory for Optical Spectroscopy of Nanostructures, Department of Experimental Physics, Faculty of Fundamental Problems of Technology, Wroc\l{}aw University of Science and Technology, 50-370 Wroc\l{}aw, Poland}
		
	\author{A.~Somers}
		\affiliation{Technische Physik, University of W\"{u}rzburg and Wilhelm-Conrad-R\"{o}ntgen-Research Center for Complex Material Systems, Am Hubland, D-97074 W\"{u}rzburg, Germany}
		
	\author{J.~P.~Reithmaier}
		\affiliation{Institute of Nanostructure Technologies and Analytics (INA), CINSaT, University of Kassel, 34132 Kassel, Germany}
		
	\author{S.~H\"{o}fling}
		\affiliation{Technische Physik, University of W\"{u}rzburg and Wilhelm-Conrad-R\"{o}ntgen-Research Center for Complex Material Systems, Am Hubland, D-97074 W\"{u}rzburg, Germany}
		\affiliation{SUPA, School of Physics and Astronomy, University of St.\,Andrews, North Haugh, KY16 9SS St.\,Andrews, United Kingdom}
		
	\author{G.~S\k{e}k}
		\affiliation{Laboratory for Optical Spectroscopy of Nanostructures, Department of Experimental Physics, Faculty of Fundamental Problems of Technology, Wroc\l{}aw University of Science and Technology, 50-370 Wroc\l{}aw, Poland}
		
	
	\begin{abstract}
		We present experimental and theoretical investigation of exciton recombination dynamics and the related polarization of emission in highly in-plane asymmetric nanostructures. Considering general asymmetry- and size-driven effects, we illustrate them with a detailed analysis of InAs/AlGaInAs/InP elongated quantum dots. These offer a widely varied confinement characteristics tuned by size and geometry that are tailored during the growth process, which leads to emission in the application-relevant spectral range of 1.25-1.65~\si{\micro\metre}. By exploring the interplay of the very shallow hole confining potential and widely varying structural asymmetry, we show that a transition from the strong through intermediate to even weak confinement regime is possible in nanostructures of this kind. This has a significant impact on exciton recombination dynamics and the polarization of emission, which are shown to depend not only on details of the calculated excitonic states but also on excitation conditions in the photoluminescence experiments. We estimate the impact of the latter and propose a way to determine the intrinsic polarization-dependent exciton light-matter coupling based on kinetic characteristics.
	\end{abstract}
	
	\maketitle
	
	\section{Introduction}\label{sec:intro}
		Extensive research has been conducted concerning the properties of correlated electron-hole pairs (excitons) in quantum dots (QDs) grown in InAs on InP material system (InAs/InP), i.e. with various barrier alloys that are lattice-matched to InP.\cite{KhanPQE2014} In spite of many similarities to self-assembled structures grown in GaAs matrix, part of the properties may vary significantly, but this has not been fully described yet. The growth kinetics in molecular beam epitaxy of InAs QDs following the self-organization process is different in case of GaAs- and InP-based structures owing to the significant inequality in lattice constant mismatch values of \SI{\sim7}{\percent}, and \SI{\sim3}{\percent}, respectively. As a result, InAs/InP QDs grown in lower strain conditions on the main crystallographic directions are of high volume and strongly elongated,\cite{RudnoRudzinskiAPL2006,SauerwaldAPL2005} in contrast to their self-assembled InAs/GaAs counterparts\cite{GoldsteinAPL1985} that are typically relatively small in all three spatial dimensions. The bigger and varying size of considered nanostructures arrives with the possibility of studying the transition of confinement regimes for excitons,\cite{EfrosFTT1982} whereas their in-plane asymmetry and additional shape imperfections allow to consider the results of confinement symmetry lowering.\cite{BayerPRB2002} Moreover, the band gap discontinuities also distinguish the InAs/InP heterostructures from the other. While the most studied InAs/GaAs material system provides high confinement potential for electrons and shallower for holes, various band offsets can be obtained with InAs on InP materials,\cite{VurgaftmanJAP2001} from strong holes' confinement in InAs/InP QDs to significantly weaker in case of InAs/AlGaInAs/InP materials' combination, as studied here. Such diversity additionally widens the possibility of exploration of various confinement regimes for excitons in quasi-zero dimensional objects, that, among other, influence especially the electron-hole Coulomb correlations.\cite{LeePRB2001} The latter is reflected in modified physical parameters, mostly the optical transition dipole moment and the resultant polarization of emitted photons as well as the exciton oscillator strength, and hence the radiative lifetime\cite{ThranhardtPRB2002}, which are the main subjects of this study.
		
		Apart from being interesting for basic research, the investigated nanostructures, thanks to their band structure properties, are utilized in quantum devices devoted to telecom applications.\cite{KhanPQE2014,LelargeJSQE2007,ReithmaierIEEE2007} Lasers, optical amplifiers, superluminescent diodes benefit from the 0D-like confinement of electrons and holes in a dot providing e.g. the threshold current insensitive to temperature, high modal gain, low sensitivity to optical feedback, broad gain and spectral tunability, all in the spectral range suitable for the second and third windows of silica fibers centered at $\SI{1.3}{\micro\meter}$ and $\SI{1.55}{\micro\meter}$, respectively.\cite{KhanPQE2014,LelargeJSQE2007,ReithmaierIEEE2007} It has also been demonstrated that a single InAs/InP QD can be used in low-loss quantum-secured data transmission lines, opening new possibilities in data processing facilitating the quantum state of light and matter.\cite{TakemotoJAP2007,BirowosutoSR2012,BenyoucefAPL2013,DusanowskiAPL2014} In all these applications the overall device functionality is strongly dependent on the parameters of Coulomb correlated electron-hole pairs (excitons) and their recombination kinetics, which are, however, not fully understood in InAs/InP 0D-like nanostructures.
	
		In this work, we focus on the temporal evolution of the exciton recombination, the associated emission process and polarization of emitted light in application-relevant InAs/Al$_{0.24}$Ga$_{0.23}$In$_{0.53}$As/InP(001) nanostructures of a widely varied confinement and emission characteristics that may be tuned by size and geometry controlled within the growth process.\cite{SauerwaldAPL2005} A number of important properties has already been revealed by structural and optical studies,\cite{SauerwaldAPL2005,MarynskiJAP2013,KhanPQE2014} which is enriched here with an extensive theoretical analysis of excitons' confinement characteristics as well as the data provided by a combination of polarization-resolved and time-resolved photoluminescence (TRPL) spectroscopies.
		
		Our theoretical considerations are based both on comprehensive reasoning drawn within simple approximations as well as on detailed and accurate modeling of the system in question followed by calculation of exciton states and their properties. Taking into account the available morphological data, we model a number of structures of various geometry and perform a detailed study of the morphology's impact on both single-particle and exciton states as well as the resulting optical properties. Then, we show that the temporal evolution of the system plays significant role in the luminescence process even under the continuous wave (CW) pumping, where a resulting excitation power dependence of polarization is present. In the case of pulsed excitation, we find a specific form of biexponential luminescence decay resulting from confinement asymmetry and its consequences in exciton states' optical properties. These considerations not only support the experimental data, but essentially deliver methods of its interpretation and proper extraction of the relevant information.
		
		Additionally, we show that the dispersion of exciton lifetime in InAs/AlGaInAs asymmetric QDs differs from that typically observed in smaller and more symmetric nanostructures.\cite{DalgarnoPRB2008,GongAPL2011} It has to be analyzed in terms of two distinct lifetimes for exciton bright states, as the two differ substantially owing to the assymetry-induced light hole admixture to the hole ground state combined with the anisotropic electron-hole exchange interaction. In addition, both lifetimes are strongly driven by the increased (and varying) role of the Coulomb correlation within exciton, which results from specific electronic structure defined by the size and asymmetry of the objects and specifically weak hole confinement.
	
		The paper is organized in the following way. In Sec.~\ref{sec:experiment} we describe the investigated system and details of the experiment along with its main results. Next, in Sec.~\ref{sec:theory} we present both approximate qualitative and accurate numerical theoretical description of the system along with resulting reasoning on ways of treatment of experimental data. Then, in Sec.~\ref{sec:results} we confront the theory with results of our measurements and present a proposition that leads to a reasonable agreement, as well as explain possible sources of experimental features not covered explicitly by the theory. Finally, we conclude the paper in Sec.~\ref{sec:conclusions}.
		
	\section{Experimental}\label{sec:experiment}
		In this section we first define the investigated system (samples) including the growth process and available morphological data, then describe the experimental setup as well as measurement techniques used, and finally present the obtained results along with explanation of their theoretically-based treatment and interpretation.
		\subsection{Structures and experimental methodology}
			The investigated nanostructures were grown in a gas source molecular beam epitaxy reactor on a sulfur-doped InP(001) substrate.\cite{SauerwaldAPL2005} Four samples were selected for the study (labeled S1-S4), each of them consisting of a \SI{200}{\nano\meter} thick Al$_{0.24}$Ga$_{0.23}$In$_{0.53}$As barrier (lattice-matched to InP) on which InAs was deposited with a varying nominal thickness: $d_\mathrm{InAs}=\SI{0.62}{\nano\meter}$ (S1), \SI{0.85}{\nano\meter} (S2), \SI{1.03}{\nano\meter} (S3), \SI{1.26}{\nano\meter} (S4). The Stranski-Krastanov growth process resulted in the formation of 0D-like nanostructures with an areal density above \SI{e10}{\centi\meter^{-2}} on a thin InAs wetting layer (WL). Finally, the dot-like structures were covered with a \SI{100}{\nano\meter} thick Al$_{0.24}$Ga$_{0.23}$In$_{0.53}$As barrier to keep them optically active and with an additional \SI{20}{\nano\meter} of InP to prevent oxidation. Structural data for uncapped objects revealed that they are preferentially aligned and elongated along the $[1\bar{1}0]$ crystallographic direction and have well-defined triangle-like cross sections with a fixed width to height ratio, $W/H\approx6$.\cite{SauerwaldAPL2005} The length ($L$) of the structures, which can vary considerably within the ensemble, is of high uncertainty. A typical structure geometry is presented in the right inset of Fig.~\ref{fig:experiment}(a). While the information on the cross-section is obtained for buried structures, the length is estimated based on scanning electron microscopy images of uncovered ones indicating a significant length variation even within the same ensemble.\cite{RudnoRudzinskiAPL2006,MusialPRB2012} It is also known that deposition of the top layer of barrier material may lead to changes in morphology, as the newly created material interface relaxes. Therefore, the possible range of $L$ values may extend from \SI{\sim 25}{\nano\metre} (similar to typical in-plane symmetric QDs) to hundreds of nanometers, which gives long and highly asymmetric objects, sometimes called quantum dashes (QDashes). All these constitute an interesting landscape of unusual confinement geometries.
			
			For all experiments the samples were kept in a continuous-flow liquid-helium cryostat at $T=\SI{4.2}{\kelvin}$ and photoexcited through a microscope objective (NA = 0.4, laser spot size $d\approx \SI{2}{\micro\meter}$) either by a \SI{660}{\nano\meter} line of a continuous-wave semiconductor diode laser or a train of laser pulses generated by a Ti:Sapphire oscillator. For the latter, the pulse width, pulse-to-pulse distance, and pulse photon wavelength were: \SI{\sim2}{\pico\second}, \SI{13.16}{\nano\second}, and \SI{826}{\nano\meter}, respectively (non-resonant excitation above the InP band gap). Radiation emitted from a sample was collected by the same objective and directed to a spectral analyzer consisting of a \SI{0.3}{\metre}-focal length monochromator and an GaInAs-based multichannel detector in the CW experiments. The degree of linear polarization of emission (DOP$_\mathrm{PL}$) was measured by rotating a half-wave plate inserted in front of a linear polarizer placed at the monochromator entrance slit. For the TRPL experiments one of the monochromator's outputs was equipped with a state-of-the-art nitrogen-cooled streak camera system based on GaInAsP photocathode from Hamamatsu with an effective temporal resolution of \SI{\sim80}{\pico\second}.
		
		\subsection{Experimental results}
			\begin{figure}[!tbp]
				\begin{center}
					\includegraphics[width=\columnwidth]{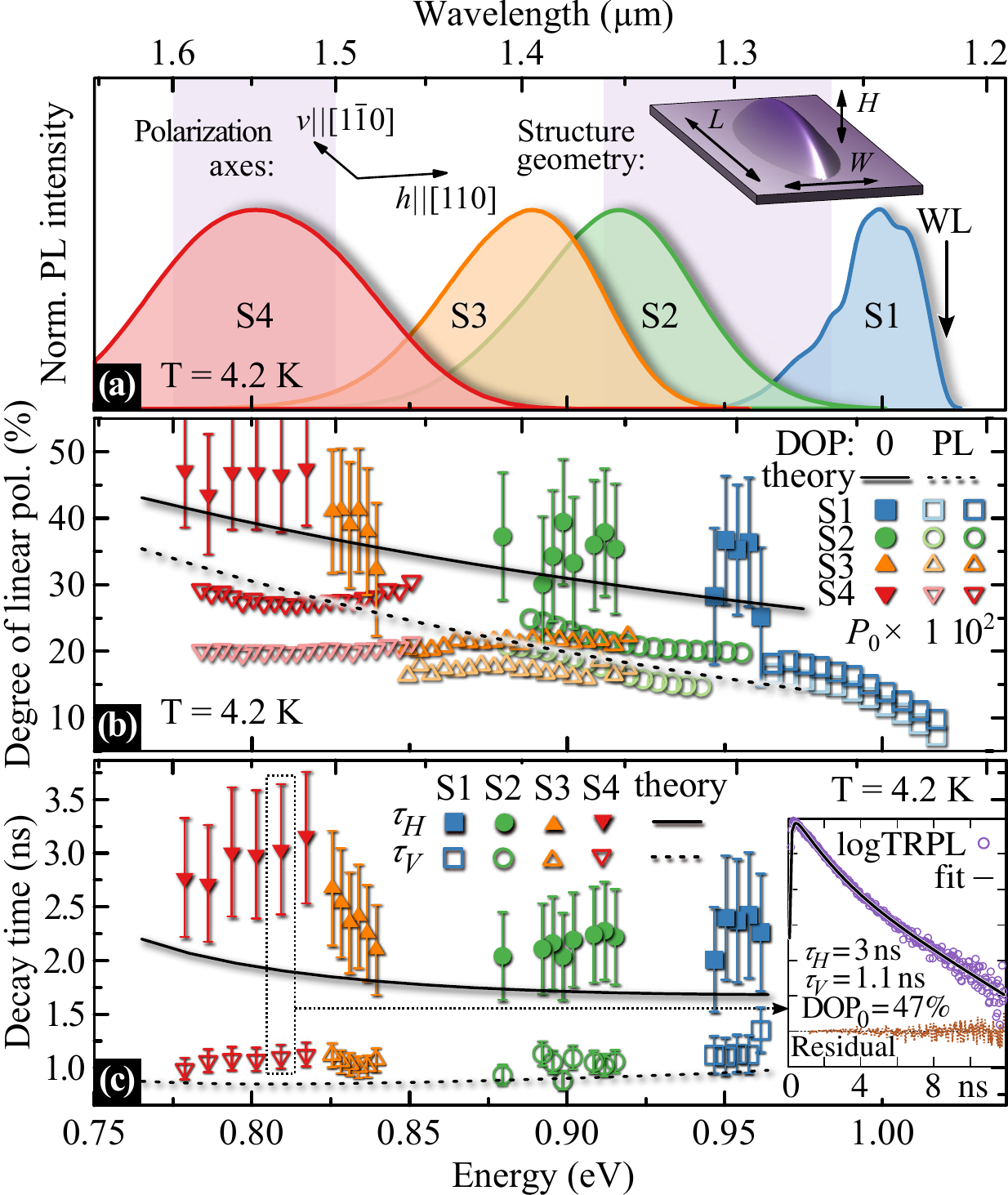}\upemhf
				\end{center}
				\caption{\label{fig:experiment}(Color online) (a) Low-temperature PL spectra obtained under non-resonant CW excitation of samples S1-S4. Approximate position of wetting layer emission marked with an arrow. Second and third optical fiber transmission windows marked with shaded areas. Insets: definition of polarization axes and a schematic view of an exemplary investigated nanostructure. (b) Dispersion of the degree of linear polarization of PL emission (DOP$_\mathrm{PL}$, open symbols) for two excitation powers (lighter symbols for $P_0=\SI{5}{\micro\watt}$, darker for $10^2 P_0$) and intrinsic DOP$_0$ (filled symbols) extracted from the time-resolved data. Black solid line marks the theoretical result for DOP$_0$, black dashed line represents the same theory with power dependence of linear polarization imposed to reproduce DOP$_\mathrm{PL}$ (see Sec.~\ref{sec:theory}). (c) Dispersion of the two decay times present in the time-resolved photoluminescence (points), and their theoretically predicted values (lines). Inset: exemplary PL decay (violet circles) for S4 at \SI{1530}{\nano\metre} with a theoretical fit with Eq.~\eqref{eq:decay_relax} (solid black line) and resulting residuals (difference of data and fit) plot (dotted brown line). }
			\end{figure}
			Low-temperature PL spectra for all the investigated samples are presented in Fig.~\ref{fig:experiment}(a). A broad spectral range of about \SIrange[]{0.75}{1.05}{\electronvolt} is covered thanks to the tuned amount of the deposited InAs material (nanostructure size). Large broadening of the PL peaks (from \SI{\sim60}{\milli\electronvolt} to \SI{\sim90}{\milli\electronvolt}) reflects the ensemble's inhomogeneity with respect to chemical content, strain, but mostly size of nanostructures. Hence, the selected set of samples can be utilized to study the impact of the varying confinement regime and structural asymmetry on exciton properties by evaluation of dispersion of the exciton-related parameters: the degree of linear polarization of emission and the exciton lifetime.
	
			The degree of linear polarization of photoluminescence,
			\begin{equation}
				\mathrm{DOP_{PL}}=\frac{I_{v}-I_{h}}{I_{v}+I_{h}}\times \SI{100}{\percent},
			\end{equation}
			is defined as the ratio of intensities of two linearly polarized components of emission, $I_{v}$ and $I_{h}$, measured with respect to optically distinguished orthogonal in-plane axes, $v\parallel[1\bar{1}0]$ and $h\parallel[1{1}0]$ (see the left inset in Fig.~\ref{fig:experiment}(a)). The polarization-resolved photoluminescence experiment allowed us to plot the dispersion of DOP$_\mathrm{PL}$ over the entire spectral range as well as within individual emission bands, as presented by open symbols in Fig.~\ref{fig:experiment}(b) for two values of excitation power. One may notice a strong trend from \SI{\sim10}{\percent} for S1 to \SI{\sim30}{\percent} in the case of S4, as well as a power dependence (lighter symbols are for lower excitation power), that for S1 is weaker than in case of other structures. The weakly polarized emission in the higher-energetic part of S1 band is possibly a result of emission from more in-plane symmetric confining potentials (possibly WL width fluctuations) rather than strongly in-plane elongated structures. This may be understood considering $d_\mathrm{InAs}$ for this sample being very close to the critical thickness between the 2D and 3D growth modes.\cite{PonchetSSE1996} This causes an overlap of the WL emission (found at approximately \SI{1.02}{\electronvolt} for structures with no QD-WL overlap\cite{RudnoRudzinskiAPL2005,RudnoRudzinskiAPL2006}, here possibly red-shifted due to partial hybridization of WL and QD states) with the emission from the actual nanostructures (low energy side of the PL peak). For the latter the DOP$_\mathrm{PL}$ is already elevated as for the rest of samples suggesting the existence of intrinsic asymmetry affecting exciton's confinement caused by non-uniform strain field, atomistic disorder at interfaces, the local asymmetry of the InAs zinc-blende crystal lattice, but mostly by the shape of nanostructures.\cite{BayerPRB2002,MusialPRB2012,ZielinskiPRB2015} Apparent dominance of $I_{v}$ over $I_{h}$ may suggest contribution of two exciton states with unequal oscillator strengths to the overall emission. Initial studies of this issue have been presented recently.\cite{TaharaPRB2013,SyperekAPL2016} One may expect this property to be reflected also in the observed exciton recombination, which should provide additional information about the exciton's confinement in the investigated nanostructures.

			In the TRPL experiment, the non-resonant pulse excitation at low average power density leads to occupation of the lowest energy exciton states, where the recombination dynamics leads to biexponential PL decays (as discussed in detail in Sec.~\ref{sec:theory}), similar to that presented in the inset of Fig.~\ref{fig:experiment}(c). These have been fitted with a specific form of biexponential function with an additional term accounting for an initial rise of the signal due to the post-pulse exciton relaxation, Eq.~\eqref{eq:decay_relax}, predicted by our theoretical considerations. This enabled us to extract two characteristic decay times, $\tau_{V}$ and $\tau_{H}$, which would be unresolvable in general due to relatively small difference in their values and similar amplitudes of both components.\cite{GrinwaldAnalBiochem1974} The corresponding decay times are plotted in Fig.~\ref{fig:experiment}(c) as a function of photon energy. One may notice a relatively strong dispersion of $\tau_{H}$, while a much weaker opposite trend is present in the case of $\tau_{V}$. The dispersive character and values of the presented data may be explained in terms of exciton's confinement conditions defined by the size and asymmetry of the studied nanostructures. This matter is addressed in Sec.~\ref{sec:theory}, where we also define the \textit{intrinsic} DOP (DOP$_0$) for exciton ground state, which essentially characterizes exciton optical properties unaffected by any experimental conditions. According to our theoretical considerations, DOP$_0$ may be extracted from the TRPL data and the corresponding values are presented in Fig.~\ref{fig:experiment}(b) with filled symbols.
			
			Our theoretical treatment of experimentally obtained PL decay profiles brought exciton radiative lifetimes different from that typically reported for similar in-plane elongated QDs in the considered spectral range.\cite{SyperekAPL2013,DusanowskiAPL2013} In view of our study, the previously reported values represent an average exciton lifetime resulting from joint treatment of the two emitting states, $\ket{H}$ and $\ket{V}$, with longer and shorter lifetimes, respectively. Moreover, recent experimental results indicated that, indeed, two different lifetimes may be present\cite{LangbeinPRB2004,TaharaPRB2013} with $\tau_V$ as short as 1.1~ns,\cite{SyperekAPL2016} and longer $\tau_H$, with discussion that the effect may be attributed to strain.\cite{TaharaPRB2013} Calculation of ``effective'' lifetimes that would correspond to an averaged oscillator strength for the two states, $\tau_\mathrm{av}^{-1}=(\tau_\mathrm{V}^{-1}+\tau_\mathrm{H}^{-1})/2$, yields \SIrange{1.45}{1.8}{\nano\second} for our experimental data, which is comparable to the previously reported values.
			
			Finally, we were able to estimate the post-pulse average exciton relaxation time to be below the limit of \SI{80}{\pico\second} defined by the temporal resolution of our experimental system, as the signal growth times were \SI[separate-uncertainty]{78\pm11}{\pico\second}, \SI[separate-uncertainty]{80\pm12}{\pico\second}, and \SI[separate-uncertainty]{80\pm14}{\pico\second} for samples S2-S4, respectively. In the case of sample S1 the initial buildup of PL signal was more complicated, possibly due to effective WL-QD occupation transfer,\cite{SyperekPRB2013} which is out of the scope of this paper and requires a separate study. Due to this, fitting of PL decays from this sample was done on cropped data which led to larger uncertainties.
		
	\section{Theory}\label{sec:theory}
		We begin this section with simple qualitative considerations on aspects of in-plane elongated QDs that are expected to have pronounced impact on recombination dynamics and optical properties of the exciton states. Next, we present a detailed and realistic modeling of the investigated nanostructures, calculation of single-particle and exciton states and their oscillator strengths describing their coupling to light, followed by evaluation of our initial predictions. Finally, we consider occupation evolution and the resulting emission characteristics for two types of optical excitation conditions, with reference to performed experiments.
		
		\subsection{Qualitative considerations}\label{subsec:model}
			The in-plane elongation of the studied QDs has a twofold impact on exciton eigenstates and their optical properties: via the resulting confinement size and the symmetry breaking.
			
			In the presence of confinement anisotropy, the nominally heavy-hole ($\ket{\Uparrow{\sla}\Downarrow}$, hh) ground state gains a light-hole ($\ket{\uparrow{\sla}\downarrow}$, lh) admixture with opposite spin,\cite{KoudinovPRB2004,LegerPRB2007} i.e. $\ket{\Uparrow'{\sla}\Downarrow'}\propto \ket{\Uparrow{\sla}\Downarrow} \pm \ii\varepsilon \ket{\downarrow{\sla}\uparrow}$, where $\varepsilon \in \mathbb{R}_{+}$ is the admixture magnitude, $\varepsilon\ll1$. Bright excitons involving a hole in such a state, $\ket{\uparrow\Downarrow'}$ and $\ket{\downarrow\Uparrow'}$ (first arrow denotes the electron and second the hole spin projection), couple to elliptically polarized light, with major polarization axes inclined towards $[1\bar{1}0]$ ($v$-axis) for both states. This makes the $v$-axis an optically preferred one, as compared to $h\parallel[110]$, and has been used to explain the nonzero DOP$_\mathrm{PL}$ from in-plane elongated QDs.\cite{MusialPRB2012,KaczmarkiewiczSST2012} The above-defined eigenstates are indeed valid for excitons in QDs with perfect cylindrical symmetry, but in the presence of the anisotropic electron-hole exchange interaction (resulting from underlying crystal lattice as well as confinement asymmetry),\cite{BayerPRB2002,ZielinskiPRB2015}
			\begin{equation*}
				H_\mathrm{ex}^{(\mathrm{eh})} = -\sum_{\mathclap{i=v,h,z}} \left(a_i J_i s_i + b_i J_i^3 s_i \right),
			\end{equation*} 
			where $s_i$ and $J_i$ are spin operator components for electrons and holes, respectively, new eigenstates are produced: dark $\ket{ {D}_{Z}/{D}_{V} }\propto \ket{\uparrow\Uparrow'}\mp\ket{\downarrow\Downarrow'}$, and bright $\ket{V/H}\propto \ket{\downarrow\Uparrow'}\mp \ii \ket{\uparrow\Downarrow'}$ ones. The latter two, in the absence of the lh admixture, couple to light polarized linearly along $v$ and $h$ axes, respectively, with equal strengths. The lh admixture results, however, in an inequality of oscillator strengths of the two states, $f_V>f_H$, which is another manifestation of $v$-axis as the optically preferred one. Calculation of optical transition dipole moments
			\begin{equation*}
				\sqrt{2}\bm{d}_{H/V} = \bm{d}_{\downarrow\Uparrow} \pm \bm{d}_{\uparrow\Downarrow} +\ii\varepsilon \bm{d}_{\downarrow\downarrow} \pm \varepsilon \bm{d}_{\uparrow\uparrow},
			\end{equation*}
			where $\bm{d}_{\downarrow\Uparrow{/}\uparrow\Downarrow}=\sqrt{3}\bm{d}_{\uparrow\uparrow{/}\downarrow\downarrow}=\frac{d_0}{\sqrt{2}}\left(\pm1,\ii,0\right)$ for zinc-blende crystals,\cite{HaugBook2004} yields the ratio of oscillator strengths
			\begin{equation*}
				\frac{f_{V}}{f_{H}}=\frac{\left|\bm{d}_{V}\right|^2}{\left|\bm{d}_{H}\right|^2}=\left(\frac{\sqrt{3}+\varepsilon}{\sqrt{3}-\varepsilon}\right)^2\approx 1 + \frac{4\sqrt{3}}{3}\varepsilon.
			\end{equation*}
			This relation should have an impact on both the exciton recombination dynamics and the polarization of emission. In the context of the latter, we define the \textit{intrinsic} DOP (DOP$_0$),
			\begin{equation}
				\mathrm{DOP}_0=\frac{f_{V}-f_{H}}{f_{V}+f_{H}}\times \SI{100}{\percent} \approx \frac{2\sqrt{3}}{3}\varepsilon \times \SI{100}{\percent},
			\end{equation}
			as a quantity characterizing the intrinsic anisotropy of the exciton-light coupling with respect to polarization. For dark excitons one has
			\begin{equation*}
				\sqrt{2}\bm{d}_{D_V\!/\!D_Z} = \bm{d}_{\uparrow\Uparrow} \pm \bm{d}_{\downarrow\Downarrow} +\ii\varepsilon \bm{d}_{\uparrow\downarrow} \mp \ii\varepsilon \bm{d}_{\downarrow\uparrow},
			\end{equation*}
			where $\bm{d}_{\downarrow\uparrow}=\bm{d}_{\uparrow\downarrow}=\frac{\sqrt{2}d_0}{\sqrt{3}}\left(0,0,1\right)$ and $\bm{d}_{\uparrow\Uparrow}=\bm{d}_{\downarrow\Downarrow}=0$,\cite{HaugBook2004} so
			\begin{equation*}
				\bm{d}_{D_V} = 0,\quad \bm{d}_{D_Z} = \ii\varepsilon \frac{2\sqrt{6} d_0}{3}\left(0,0,1\right),
			\end{equation*}
			and a brightening of $\ket{D_Z}$ is present with a significant oscillator strength $f_{D_Z}\approx4\varepsilon^2f_{V\!H}/3$, where $f_{V\!H}=(f_V+f_H)/2$. However, the resulting emission is linearly polarized along the $z$-axis, i.e. propagates parallel to the sample plane, so it is not measured in typical optical experiments.

			Another relevant consequence of the in-plane QD elongation is a modification of energy ladders for electrons and holes. Typically, in self-assembled QDs, one deals with relatively small anisotropy splitting between $p$-type states that are separated from the ground state by several tens of meV or more. Employing a simple harmonic confinement model to elongated QDs yields an energy ladder scaling like $E_{ijk}\propto i L^{-2}+j W^{-2}+k H^{-2}$, where $i$, $j$, $k$ number excitations along the respective QD dimensions. While the ground state energy, defined mostly by the QD height is weakly influenced by its in-plane elongation, the lowest energy level spacing, especially the $s$-$p$ splittings for electrons and holes ($\Delta_{sp}^{\mathrm{(e/h)}}$), scale within this approximation as $\eta^{-2}$, where $\eta=L/W$ is the in-plane aspect ratio. Additionally, the energy ladder of hole states is expected to be denser, not only due to higher effective mass, but also a very small valence band offset (VBO) in the investigated material system (\SI{\sim 80}{\milli\electronvolt} for an unstrained interface as calculated according to values presented in Table~\ref{tab:material}), leading to a shallow hole confining potential of about \SI{0.1}{\electronvolt} in a QD. This, combined with a typical value of the electron-hole Coulomb interaction energy in QDs, $\Delta_\mathrm{X}\sim\SIrange[range-phrase=\mhyphen,range-units=single]{10}{20}{\milli\electronvolt}$,\cite{CornetPRB2006, HolmJAP2002, XiaoJAP1999} may lead to a transition from more common strong ($\Delta_{sp}^{\mathrm{(e/h)}}\gg\Delta_\mathrm{X}$) to intermediate ($\Delta_{sp}^{\mathrm{(e)}}>\Delta_\mathrm{X}>\Delta_{sp}^{\mathrm{(h)}}$) or even weak ($\Delta_{sp}^{\mathrm{(e/h)}}\ll\Delta_\mathrm{X}$) confinement regime.\cite{EfrosFTT1982} For the two latter ones, when $\Delta_\mathrm{X}$ becomes at least comparable with the level spacing, one deals with a significant contribution of higher electron and/or hole states to the exciton ground state, which leads to the increase of its oscillator strength. This purely quantum effect may be considered as the ability of an exciton to recombine independently through each of the channels brought by its components. Let us consider two lowest electron and hole states, $\smket{\mathrm{e}_{1(2)}}$, $\smket{\mathrm{h}_{1(2)}}$, and an exciton in a superposition $\ket{X}=\left(\ket{\mathrm{e}_1 \mathrm{h}_1}+\epsilon \ket{\mathrm{e}_2 \mathrm{h}_2}\right)/\sqrt{1+\epsilon^2}$, where $\epsilon\in\mathbb{R}_{+}$ for simplicity. Assuming fully overlapping electron and hole envelopes and the two states to be equally bright (dipole moments $\bm{d}$), one obtains
			\begin{equation}
				\left\lvert \bm{d}_X \right\rvert^2=\frac{\left\lvert \bm{d} + \epsilon \bm{d} \right\rvert^2}{1+\epsilon^2} = \left\lvert \bm{d} \right\rvert^2 \left(1+\frac{2\epsilon}{1+\epsilon^2}\right) > \left\lvert \bm{d} \right\rvert^2,
			\end{equation}
			which gives $2\left|\bm{d}\right|^2$ for an equal superposition ($\epsilon=1$). This exaggerated example evidently lacks taking into account that not all states are equally bright and dipole moments may add up destructively. We will, however, show that even a relatively small contribution of higher states actually leads to a significant rise of the oscillator strength. 
			
			Additionally, the character of the first few excited states changes. As excitations along the QD elongation should be less energetic, one may expect the $p$-type state oriented across the QD (i.e. along the $h$-axis, labeled $p_h$; see Fig.~\ref{fig:states}(c)) to be preceded by a number of excited states with envelope nodes oriented along the QD ($\lfloor \eta^2 \rfloor$ in the simple box or harmonic confinement approximation).
		
		\subsection{Numerical modeling}\label{subsec:modeling}
			\begin{figure}[!tbp]
				\begin{center}
					\includegraphics[width=\columnwidth]{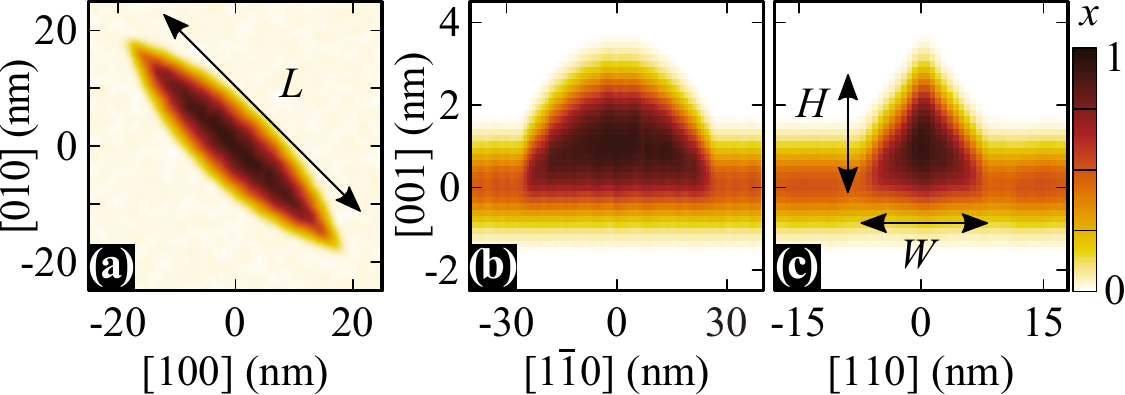}\upemhf
				\end{center}
				\caption{\label{fig:structure}(Color online) An exemplary simulated structure material composition (Al$_{0.24-0.24x}$Ga$_{0.23-0.23x}$In$_{0.53+0.47x}$As) cross-sections in planes normal to (a) $[001]$ (b) $[110]$ and (c) $[1\bar{1}0]$ directions with quantum dot dimensions marked.}
			\end{figure}
			
			\begingroup
			\begin{table}[!tbp]
				\newcommand{\cw}{\cite{WinklerBook2003}}
				\newcommand{\cs}{\cite{SaidiJAP2010}}
				\newcommand{\ct}{\cite{TseJAP2013}}
				\newcommand{\ca}{\cite{AmirtharajBOOK1994}}
				\newcommand{\cmsq}{${\sfrac{\mathrm{C\!}}{\mathrm{m}^2}}$}
				\newcommand{\intEg}{$^{-0.13}_{+1.31x}$}
				\begin{tabular*}{\columnwidth}{l|@{\extracolsep{\fill}}cccccc}
					\toprule\rule{0pt}{1.1em} 
					\phantom{aaa}& ~~~AlAs~~~ & ~~~GaAs~~~ & ~~~InAs~~~ & ~$C^\mathrm{AlAs}_\mathrm{GaAs}$~ & ~$C^\mathrm{GaAs}_\mathrm{InAs}$~ & ~$C^\mathrm{AlAs}_\mathrm{InAs}$ \\
					\hline\rule{0pt}{1.1em} 
					$a_\mathrm{}$\,(\AA) 			& 5.66	&5.65	&6.06	&0		&0		&0\\
					$E_\mathrm{g}$\,(eV)			& 3.1	&1.519	&0.417	&\intEg	&0.477	&0.7\\
					VBO\,(eV)						& -1.33	&-0.80	&-0.59	&0		&-0.38	&-0.64\\
					$E_\mathrm{p}$\,(eV)\cs$^{*}$ 	& 19.15	&23.8	&19.33	&0		&0		&0\\
					$m_\mathrm{e}^{*}$ 				& 0.15	&0.067	&0.026	&0		&0.0091	&0.049\\
					$\Delta_\mathrm{}$\,(eV) 		& 0.28	&0.341	&0.39	&0		&0.15	&0.15\\
					$\gamma_\mathrm{1}$$^{**}$		& 3.76	&6.98	&20.0	&0		&0		&0\\
					$\gamma_\mathrm{2}$$^{**}$		& 0.82	&2.06	&8.5		&0		&0		&0\\
					$\gamma_\mathrm{3}$$^{**}$		& 1.42	&2.93	&9.2		&0		&0		&0\\
					$e_\mathrm{14}$\,(\cmsq)\ct 		& -0.055	&-0.16	&-0.045	&0		&0		&0\\ 
					$B_\mathrm{114}$\,(\cmsq)\ct 		& -0.653	&-0.666	&-0.653	&0		&0		&0\\
					$B_\mathrm{124}$\,(\cmsq)\ct 		& -1.617	&-1.646	&-1.617	&0		&0		&0\\
					$C_\mathrm{k}$\,(eV\AA) 			& 0.002	&-0.0034	&-0.0112	&0		&0		&0\\
					$a_\mathrm{c}$\,(eV)	 		& -5.64	&-7.17	&-5.08	&0		&2.61	&-1.4\\
					$a_\mathrm{v}$\,(eV) 			& 2.47	&1.16	&1.0		&0		&0		&0\\
					$b_\mathrm{v}$\,(eV) 			& -2.3	&-2.0	&-1.8	&0		&0		&0\\
					$d_\mathrm{v}$\,(eV) 			& -3.4	&-4.8	&-3.6	&0		&0		&0\\
					$c_\mathrm{11}$\,(GPa)			& 1250	&1211	&833		&0		&0		&0\\
					$c_\mathrm{12}$\,(GPa) 			& 534	&548		&453		&0		&0		&0\\
					$c_\mathrm{44}$\,(GPa) 			& 542	&600		&396		&0		&0		&0\\
					$\varepsilon_\mathrm{r}$\cw 		&10.06	&12.4	&14.6	&0		&0		&0\\
					$n_\mathrm{}$\ca 				&2.87	&3.347	&3.42	&0		&0		&0\\
					\toprule
					\multicolumn{7}{p{\columnwidth}}{\rule{0pt}{1.em} 
						$\!^{*}$Values used for calculation of optical properties; for the $\kp$ Hamiltonian $E_{P} = 
						\left (\mfrac{m_{0}}{m_\mathrm{e}^{*}} - 1 \right ) \mfrac{E_\mathrm{g}(E_\mathrm{g}+\Delta)}{(E_\mathrm{g}+2\Delta/3)}$ was used to preserve ellipticity of the $\kp$ equation system for envelope functions\cite{VeprekPRB2007,BirnerBOOK2014}.}\\
					\multicolumn{7}{p{\columnwidth}}{
						$^{**}$Interpolation done for the inverse of $\gamma_i$.\cite{StierPRB1999}}
				\end{tabular*}\upemhf
				\caption{\label{tab:material} Material parameters used in the modeling of nanostructures and calculation of single-particle and exciton states; $C^\mathrm{A}_\mathrm{B}$ are values of ternary bowing parameters. Unless otherwise marked, parameters taken after Ref.~[\onlinecite{VurgaftmanJAP2001}], where the interpolation formula for a quaternary alloy may also be found.\upemhf}
			\end{table}
			\endgroup
			Structures in question were modeled as triangular in the cross-section with a fixed $W/H=6$ ratio, according to available morphological data,\cite{SauerwaldAPL2005} and with elliptical height profile along the elongation axis, revealed recently for similar structures via atom probe tomography (APT),\cite{MarynskiJAP2013} protruding from a \SI{0.9}{\nano\metre} thick wetting layer (WL), as illustrated in the inset in Fig.~\ref{fig:experiment}(a). The material composition profile, visible in Fig.~\ref{fig:structure} in three cross-sections, is a result of noise added to simulate material inhomogeneity as well as intermixing between InAs and the barrier material simulated by Gaussian averaging with standard deviation of \SI{0.6}{\nano\metre}, again supported by the APT data for similar structures.\cite{MarynskiJAP2013} A total of 32 structures were simulated with $H$ varying between \SI{1.8}{\nano\metre} and \SI{3.8}{\nano\metre} nm and $L$ from approximately equal to $W$ up to about \SI{100}{\nano\metre}. In-plane dimensions were read from final material profiles to take into account the fact that intermixing alters small objects relatively stronger than large ones, so using initial spatial extents would be incorrect.

			\begin{figure}[!tbp]
				\begin{center}
					\includegraphics[width=\columnwidth]{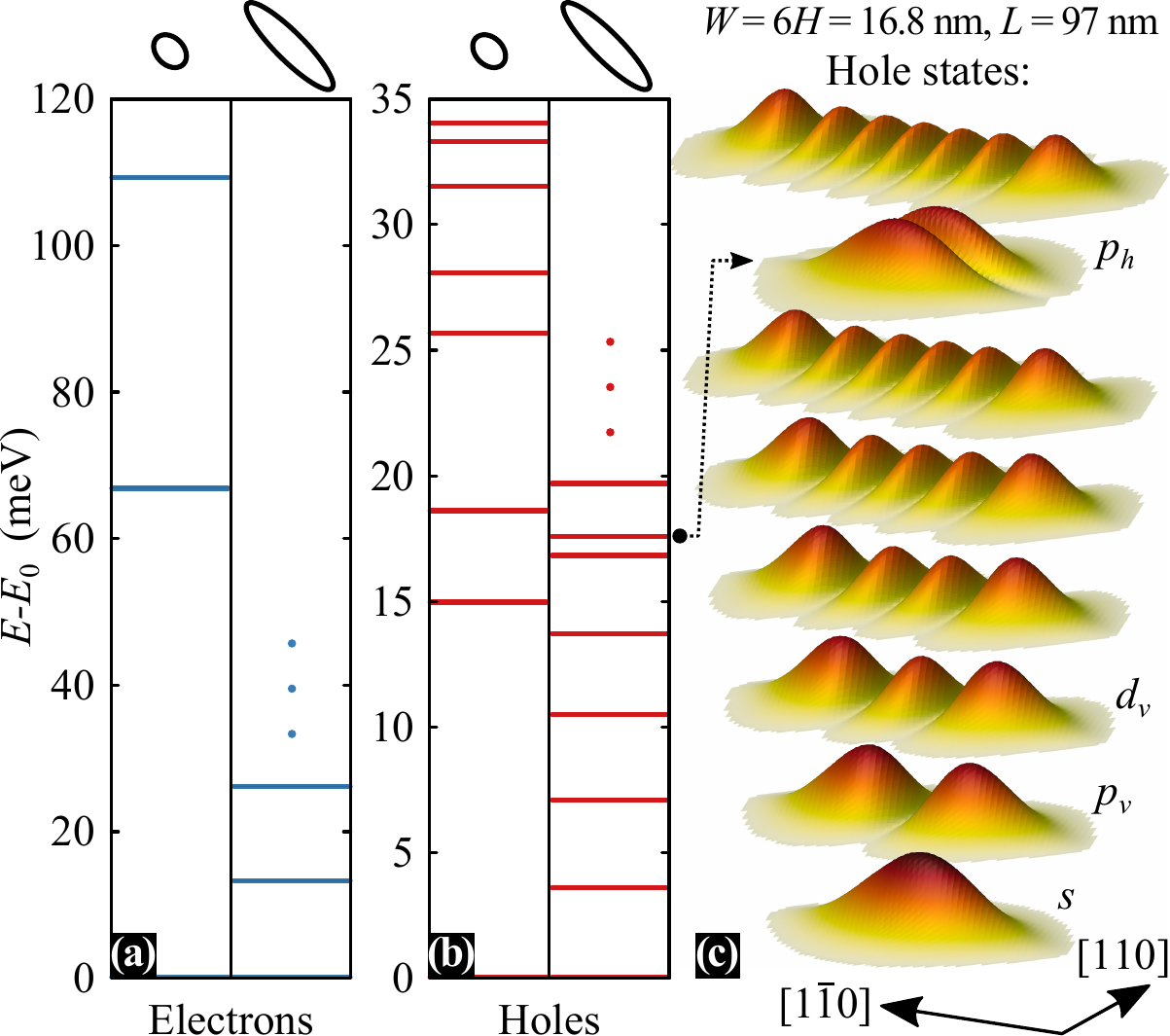}\upemhf
				\end{center}
				\caption{\label{fig:states}(Color online) Comparison of single-particle energy ladders between a nearly symmetric and a highly in-plane elongated quantum dot for (a) electrons, (b) holes; ladders for elongated dots are truncated, for sake of comparison, to present the same number of states in both cases. (c) Corresponding hole probability densities for the highly elongated dot.}
			\end{figure}
			\begin{figure}[!tbp]
				\begin{center}
					\includegraphics[width=\columnwidth]{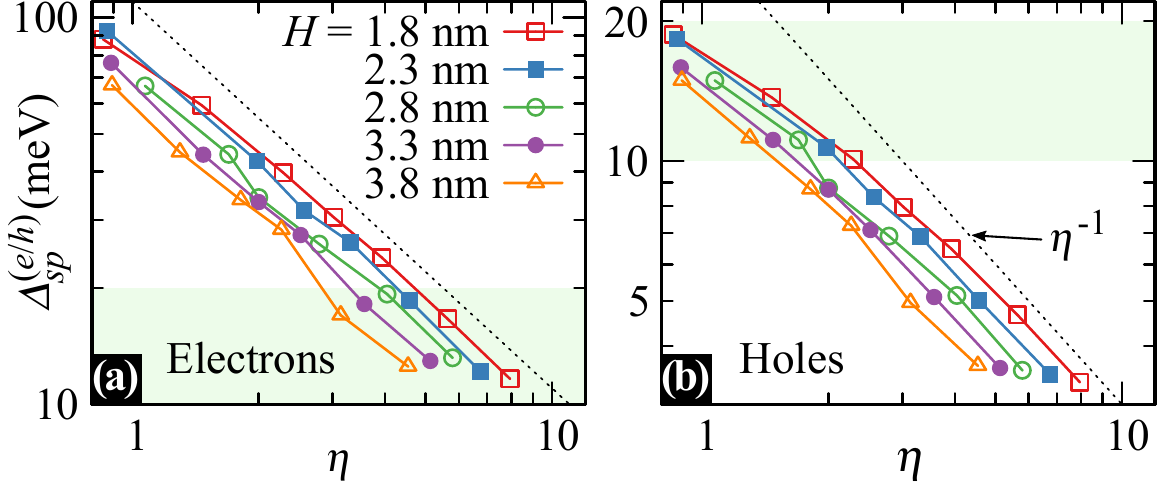}\upemhf
				\end{center}
				\caption{\label{fig:spsplit_eta}(Color online) Dependence of the $s$-$p$ splitting energy on quantum dot in-plane aspect ratio $\eta$ for (a) electrons ($\cramped{\Delta_{sp}^{\mathrm{(e)}}}$) and (b) holes ($\cramped{\Delta_{sp}^{\mathrm{(h)}}}$) for quantum dots of various height $H$. Solid lines are to guide the eye. The dotted black lines mark an $\eta^{-1}$ dependence and the shaded areas cover a typical range of electron-hole Coulomb interaction energy $\Delta_\mathrm{X}$.}
			\end{figure}
			The conduction and valence band states were calculated within the 8-band $\kp$ theory with the envelope function approximation.\cite{BurtJPhysCondMat1992,ForemanPRB1993,BahderPRB1990} The spin-orbit effects and magnetic field\cite{AndlauerPRB2008} (\SI{10}{\milli\tesla} used just to lift the Kramers' degeneracy and fix the spin basis), as well as strain\cite{BirBOOK1974,PryorPRB1998} and resulting piezoelectric field\cite{BesterPRL2006,SchulzPRB2011,TseJAP2013} up to the second order in polarization were included (see Ref.~[\onlinecite{GawareckiPRB2014}] for details of the model and numerical methods, and Table~\ref{tab:material} and references therein for material parameters used). In Fig.~\ref{fig:states}(a) and (b) we present electron and hole energy ladders for an approximately in-plane symmetric QD and a highly elongated one, both of the same height $H=\SI{2.8}{\nano\metre}$. The first characterizes with a typical electronic structure and relatively large $\Delta_{sp}^{\mathrm{(e)}}\approx\SI{66}{\milli\electronvolt}$, a few times smaller $\Delta_{sp}^{\mathrm{(h)}}$, and a moderate splitting of $p$-type states, resulting from factual asymmetry of the shape (triangular vs. elliptical cross section). In the case of the elongated QD, the ladder of excited states and their character (see Fig.~\ref{fig:states}(c) for exemplary hole probability densities) are significantly modified, as qualitatively expected. However, the $p_h$ state appears as the 6-th excited state, earlier than it might have been expected for $\eta=5.8$. In Fig.~\ref{fig:spsplit_eta} we present the dependence of ${\Delta_{sp}^{\mathrm{(e/h)}}}$ on $\eta$ against a typical range of $\Delta_\mathrm{X}$ values. One may notice the anticipated confinement regime transition, as the level spacing for electrons becomes comparable with $\Delta_X$ and for holes decreases even far below it. Interestingly, an $\eta^{-1}$ (and hence $L^{-1}$ or even weaker for holes at small values of $\eta$) dependence is evidently present, which is responsible for earlier than expected appearance of the $p_h$ state. This may be understood considering the specific shape of QDs, which cannot be approximated by a simple box or harmonic confinement model. Due to the height profile, the three main axes of confinement become dependent, as states more extended along the QD experience effectively stronger confinement in other directions. As a result, the effects of in-plane elongation are less pronounced than it might have been expected. Nevertheless, in the view of experimental evidences for such a morphology,\cite{MarynskiJAP2013} we consider it to be rather a feature of the system than an artificial theoretical imposition.

			\begin{figure}[!tbp]
				\begin{center}
					\includegraphics[width=\columnwidth]{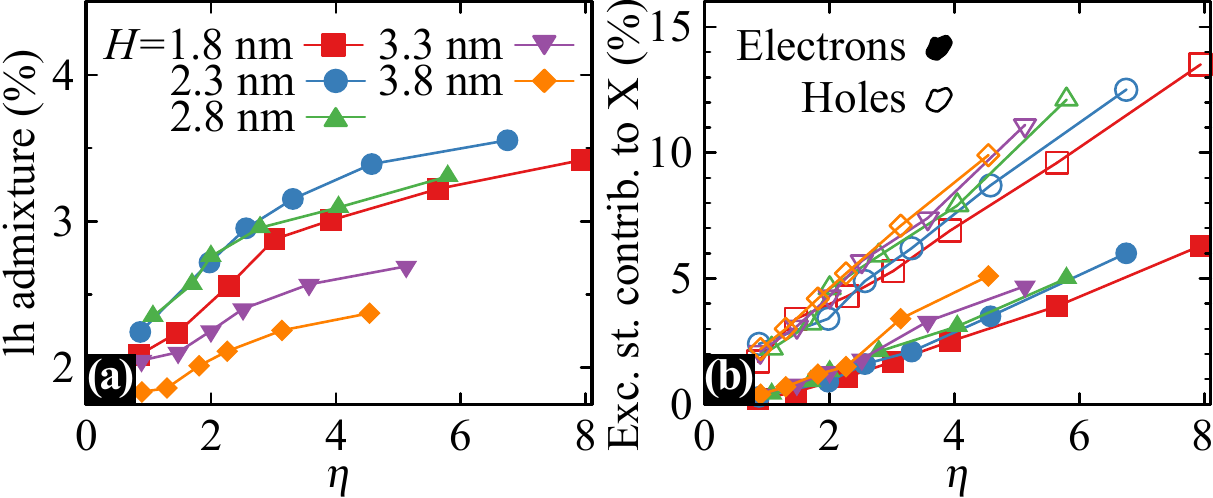}\upemhf
				\end{center}
				\caption{\label{fig:lh_admix_eta}(Color online) Quantum dot in-plane aspect ratio dependences of (a) contribution of light hole with opposite spin to the hole ground state, (b) contribution of excited electron (filled symbols) and hole (open symbols) single-particle states to the exciton ground state for quantum dots of various height $H$. Lines are to guide the eye.}
			\end{figure}
			The other anticipated effect of structural in-plane asymmetry, namely the resulting magnitude of the lh admixture is plotted as a function of $\eta$ in Fig.~\ref{fig:lh_admix_eta}(a). Besides the expected rise with increasing elongation, one may also notice a non-monotonous behavior with QD height, which we find as another effect caused by the specific shape of QDs under study.
			
			We use the basis of 6 electron and 16 hole lowest energy states to construct an excitonic basis via diagonalization of Coulomb interaction and phenomenological anisotropic electron-hole exchange interaction (with bright-dark $\Delta_0=\SI{0.4}{\milli\electronvolt}$, bright $\Delta_1=\SI{0.1}{\milli\electronvolt}$ and dark $\Delta_2=\SI{0.005}{\milli\electronvolt}$ exciton splittings based on experimental\cite{MrowinskiPRB2016} and theoretical\cite{ZielinskiPRB2014,ZielinskiPRB2015} estimations, as well as bright-dark mixing caused by a possible $C_{2v}$ symmetry breaking\cite{ZielinskiPRB2015,ZielinskiArxiv2016}), within the configuration interaction approach. In Fig.~\ref{fig:lh_admix_eta}(b) we plot the total contribution of configurations involving excited electron and hole single-particle states to the exciton ground state as a function of $\eta$. The transition between confinement regimes is visible mainly in high contribution of excited hole states, which reaches \SI{13.5}{\percent} for $\eta\approx8$.

			\begin{figure}[!tbp]
				\begin{center}
					\includegraphics[width=\columnwidth]{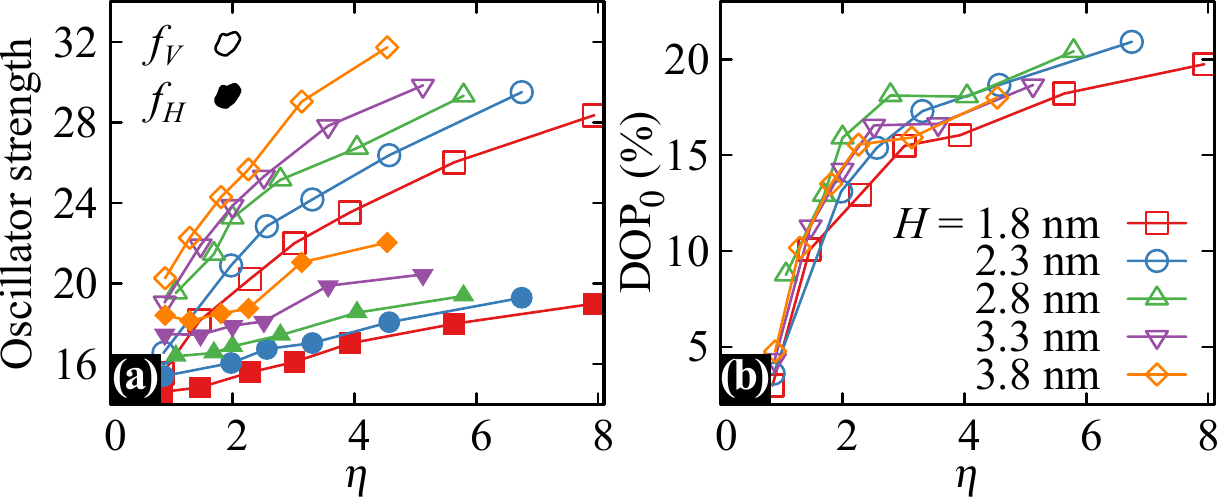}\upemhf
				\end{center}
				\caption{\label{fig:f_dop_eta}(Color online) Quantum dot in-plane aspect ratio dependence of (a) oscillator strengths for exciton states $\ket{V}$ (open symbols) and $\ket{H}$ (filled symbols) and (b) resulting intrinsic degree of polarization DOP$_0$ for quantum dots of various height $H$. Lines are to guide the eye.}
			\end{figure}
			Finally, the oscillator strength for the $i$-th calculated exciton state may be obtained as\cite{AndrzejewskiJAP2010}
			\begin{equation}
				f_i^{(j)} = {\frac{2}{m_0 E_i}\Bigg\lvert\sum_{\mathclap{\alpha,\beta}}c_{\alpha\beta}\matrixel{\bm{\psi}_\mathrm{v}^{(\alpha)}}{\hat{e}_j\cdot\bm{P}}{\bm{\psi}_\mathrm{c}^{(\beta)}}\Bigg\rvert^2},
			\end{equation}
			where $j$ labels the polarization axis, $\hat{e}_j$ is the polarization unit vector, $E_i$ is the exciton energy, $\alpha$ ($\beta$) enumerates valence (conduction) single particle states, $c_{\alpha\beta}$ are the expansion coefficients in the electron-hole configuration basis, $\bm{\psi}_{\mathrm{v(c)}}^{(\alpha(\beta))}$ is an 8-component pseudo-spinor of electron envelope functions for the $\alpha$-th valence ($\beta$-th conduction) eigenstate spanned in the standard 8-band $\bm{k}\cdot\bm{p}$ basis, $\bm{P}=\left({m_0}/{\hbar}\right)\left({\partial H_{\kp}}/{\partial \bm{k}}\right)$ is the corresponding representation of momentum operator,\cite{VoonPRB1993} and hole states are considered as time-reversed valence electrons.\cite{AndrzejewskiJAP2010} We label the four lowest states as previously, i.e. $\ket{D_Z}$, $\ket{D_V}$, $\ket{V}$, and $\ket{H}$ (in order of increasing energy) as their properties are in agreement with qualitative considerations above. Two features of dark excitons are worth noting here: a moderate $f_{D_Z}^z$ oscillator strength corresponding to $\tau_{D_Z}\sim \SI{100}{\nano\second}$ (varying with elongation), and brightening of $\ket{D_V}$ exhibited in nonzero $f_{D_V}^v$. The latter, which is the reason for the label given to this state, was absent in our qualitative calculation as we did not take into account electron-hole exchange terms arising from $C_{2v}$ symmetry breaking.\cite{ZielinskiPRB2015,ZielinskiArxiv2016} To check our preliminary predictions for bright states, we plot in Fig.~\ref{fig:f_dop_eta}(a) the dependence of $f_V$ and $f_H$ on $\eta$. One may notice the anticipated interplay between rising mismatch of the two (caused by the growing lh admixture) and the rise of both (confinement regime transition). Additionally in panel (b) the resulting DOP$_0$ is plotted.
			
			The exciton recombination rates, split into emission with each of the polarizations, $\gamma_i^{(j)}$, and the corresponding radiative lifetimes ${\tau_i^{-1}=\gamma_i=\sum_j\gamma_i^{(j)}}$ may be calculated as\cite{ThranhardtPRB2002}
			\begin{equation}
				\gamma_i^{(j)}=f_i^{(j)} E_i^2\frac{ e^2 }{6 \pi \hbar^2 c^3 \epsilon_0 m_0} \frac{9 n_\mathrm{B}^2}{2 n_\mathrm{B}^{\nicefrac{1}{2}} + n_\mathrm{QD}^{\nicefrac{1}{2}} },
			\end{equation}
			where $n_\mathrm{B}$ and $n_\mathrm{QD}$ are the bulk matrix and QD material refractive indices, respectively, $e$ is the elementary charge, and $E_i$ the exciton energy. One may notice, that the increase of the oscillator strength with the QD size may be compensated by associated decrease of exciton energy. In the case of in-plane elongated QDs, however, the first is connected mostly with the QD length, impact of which on the ground state energy is only minor. Hence, we expect the resultant lifetimes to be reduced as compared to typical QDs.
			
		\subsection{Exciton evolution: continuous wave excitation}\label{subsec:cw}
			The emission of light from nanostructures is commonly addressed theoretically with assumption that for non-resonant excitation conditions the intensity of emission from a given exciton state is directly proportional to its oscillator strength. While such a correspondence between the amount of exchanged energy and the strength of light-matter coupling is reasonable in the resonant absorption process, the emission under continuous non-resonant pumping is additionally influenced by the evolution of the system, which may be non-negligible.
			
			Non-resonantly generated photocarriers forming hot free electron and hole gases in the bulk material correlate, upon relaxation, into a gas of excitons. Further energy dissipation results in trapping of these in QDs, where, after a fast phonon-mediated relaxation, they occupy the lowest energy states. Two things need to be noted here: after a complicated series of transitions, an exciton reaches a QD with a random spin configuration, and then spin-flip mechanisms, if available, have been estimated to be inefficient.\cite{FujisawaNat2002,AtatureSci2006,KhaetskiiPRB2000,TsitsishviliPRB2003,RoszakPRB2007} Under such assumption, the system can thermalize only partially, along each of the spin-compatible sub-ladders of states (defined by the electron-hole exchange interaction), each of them obtaining equal total probability of occupation. The excitation power, which may be translated into the rate of exciton generation per a QD, $\Gamma_P$, is crucial for the integrated emission intensities. We consider for simplicity a low temperature limit and very fast relaxation, i.e. only lowest four states become occupied. In the high-power limit, $\Gamma_P\gg \gamma_i$, all occupations are constantly maximal and each of the states emits with a rate given by its oscillator strength, as it is commonly assumed to be in general. On the other hand, in the opposite limit the effective rate of emission is set by $\Gamma_P$ and equal for all four states, including slightly brightened dark ones. The transition between the two regimes is continuous and for each of the states governed by its radiative lifetime. At non-zero temperature occupation probabilities are additionally thermally distributed along spin-defined sub-ladders and corresponding intensities are weighted.

			\begin{figure}[!tbp]
				\begin{center}
					\includegraphics[width=\columnwidth]{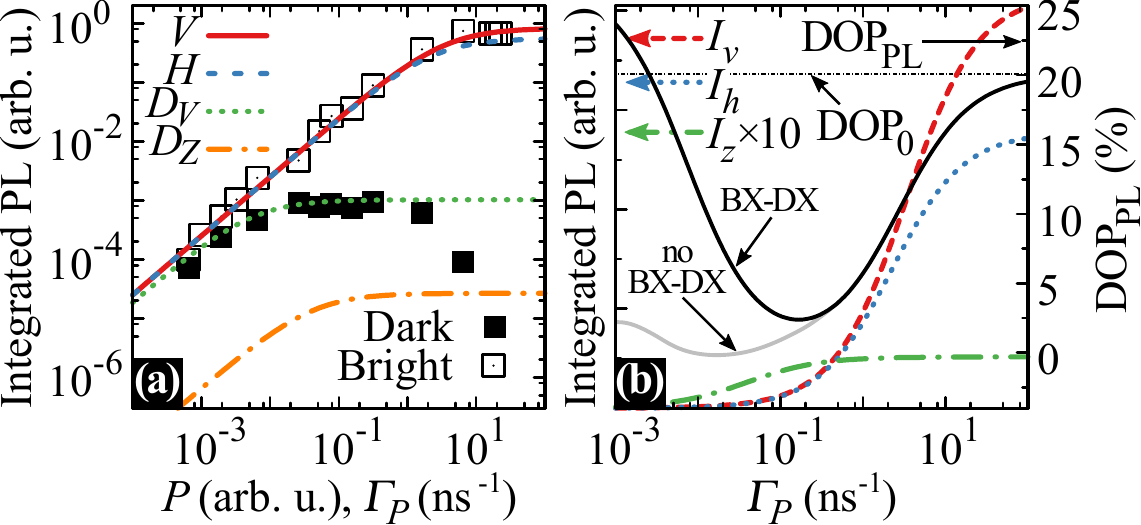}\upemhf
				\end{center}
				\caption{\label{fig:cw_power}(Color online) Effective power $\Gamma_P$, dependence of (a) emission intensities from four lowest energy exciton states calculated via Eq.~\eqref{eq:kinet_cw_lowT} (lines) compared with experimental values for bright (open symbols) and dark exciton (filled symbols) taken from Ref.~[\onlinecite{SchwartzPRX2015}], (b) total emission intensity for each of the polarization axes (left axis) and resulting PL degree of polarization DOP$_\mathrm{PL}$ (right axis) calculated for continuous wave excitation of an exemplary simulated quantum dot with and without bright-dark exciton mixing.}
			\end{figure}
			The corresponding low-temperature kinetic equations for occupation of each of the four lowest energy states, $p_i\!\left(t\right)$, take the form
			\begin{equation}\label{eq:kinet_cw_lowT}
				\dot{p}_i\!\left(t\right)=\frac{ \Gamma_{P}}{4} \left[ 1 - {p}_i\!\left(t\right) \right] -{p}_i\!\left(t\right) \sum_{\mathclap{j=v,h,z}} \gamma_i^{(j)},
			\end{equation}
			where the rates of relaxation from higher energy states are assumed to be much bigger than $\Gamma_{P}/4$. The corresponding emission intensities are $I_i^{(j)}\propto {p}_i\!\left(t\right)\gamma_i^{(j)}$, and the integrated signal in a PL experiment is proportional to the steady-state solution, $I_i^{(j)}\!\left(\infty\right)\propto{\Gamma_{P} \gamma_i^{(j)} }/{\left( \Gamma_{P} + 4 \gamma_i \right)}$. For collection of light propagating normally to the sample only $I_v$ and $I_h$ give rise to the measured signal, while a possible non-zero $I_z$ effectively weakens it, which is the case mostly for state $\ket{D_Z}$. In Fig.~\ref{fig:cw_power}(a) we present the resulting power dependence of emission intensity for the four states, where one may notice a transition from the regime of equal contribution of all states except $\ket{D_V}$ in absence of bright-dark exciton mixing (for very low power) to the emission dominated by bright states. While the latter is commonly assumed, it has to be stressed out that the typical excitation conditions ($\Gamma_P\sim \SIrange[range-phrase=\mhyphen,range-units=single]{e-2}{e2}{\nano\second^{-1}}$) span across the transition between the two regimes, which is of a big importance for polarization of emitted light. While the model used by us to obtain Eq.~\eqref{eq:kinet_cw_lowT} is very simple, it fits well the results of recent experimental measurements of power-dependent PL from bright and dark excitons\cite{SchwartzPRX2015} (marked with squares in Fig.~\ref{fig:cw_power}(a)), which we take as a confirmation of its applicability. The resulting estimated dependence of the degree of polarization of PL emission (DOP$_\mathrm{PL}$) on $\Gamma_{P}$ is shown in Fig.~\ref{fig:cw_power}(b) (solid black line) along with total intensities for each of polarization axes. A measurement of power-dependent DOP for a similar system has been recently performed\cite{MusialPRB2012} and brought an asymptotically compatible result, however, with the rise of DOP extended over more decades of power. Providing Eq.~\eqref{eq:kinet_cw_lowT} we made an assumption that $\Gamma_P/4$ is much smaller than the rate of carriers' relaxation in a QD. This may be estimated to be fulfilled in our experiment thanks to the experimentally estimated exciton relaxation times below \SI{80}{\pico\second} (see Sec.~\ref{sec:experiment}), but obviously has to break down at some point with rising power. The weakening of the rise of DOP with increasing power occurs then possibly due to an effectively sublinear relation between the rate of ground state population and power. This may result from both relaxation of carriers through the ladder of single-particle states and cascade-like kinetics of complexes involving more particles. Nonetheless, a conclusion that a high effective rate of population of the ground state is needed for excitons to exhibit DOP$_\mathrm{PL}\sim \mathrm{DOP}_0$ holds, and the effects mentioned here constitute an additional difficulty in this matter. The above-mentioned suggests, that excitonic DOP$_0$ is only partially accessible via PL measurements.
			
		\subsection{Exciton evolution: pulsed excitation}\label{subsec:pulse}
			In time-resolved experiments, results of which are presented here, each pulse generates typically less than one electron-hole pair per a QD, which at low $T$ results in statistically equal occupation of the four lowest exciton states. Each of these, having enough time after the pulse, would then emit the same amount of energy at appropriate rate $\gamma_i$. A typical repetition rate for pulsed lasers (here \SI{0.07}{\nano\second^{-1}}) is however much higher than recombination rates of dark excitons (despite the brightening of $\ket{{D_V}}$ and relatively high $\gamma_{D_Z}^z$) and leads to their CW-like behavior resulting in approximately constant low intensity emission with mostly $v$-polarized normally propagated component. This background is yet of no relevance for a TRPL experiment, where the variable emission comes equally (in terms of total emitted energy, i.e. integrated intensity) from $\ket{V}$ and $\ket{H}$ recombining at unequal rates. It is therefore reasonable to expect PL decays of the form
			\begin{equation*}\label{eq:decay}
				I\lr{t} = A \left( \gamma_{V} \mr{e}^{- \gamma_{V} t} + \gamma_{H} \mr{e}^{- \gamma_{H} t} \right) + I_0,
			\end{equation*}
			where $I_0$ is the constant background and the impact of the very weak $z$-polarized emission is neglected. The theoretically predicted connection between the relative amplitudes and rates of the two components makes the fitting procedure feasible, which is generally not the case for two exponentials with comparable rates and amplitudes.\cite{GrinwaldAnalBiochem1974} Most of the information on the faster component is contained in the initial part of the decay, which is additionally influenced by a fast relaxation from higher states leading to an initial buildup of PL signal. We find it beneficial to include this effect for the price of an additional fitting parameter rather than to exclude this crucial part of data. The relaxation is modeled by an average rate of population of bright exciton states from the set of higher ones that are assumed to be populated instantaneously at the pulse arrival time (since the pulse duration is negligible on the considered time scale in our case). This yields 
			\begin{equation}\label{eq:decay_relax}
				I\lr{t} = A \sum_{i={V},{H}} \cramped{\frac{\mr{e}^{\raisemath{0.4em}{-\frac{t-t_0}{\tau_i}}}-\mr{e}^{\raisemath{0.4em}{-\frac{t-t_0}{\tau_\mathrm{r}}}}}{\tau_i-\tau_\mathrm{r}}} + I_0,
			\end{equation}
			where $t_0$ is the pulse arrival time (fixed for properly prepared data), and $\tau_\mathrm{r}$ the effective relaxation time. Determination of $\tau_{V}$ and $\tau_{H}$ by fitting such decay curves to the TRPL data allows for experimental estimation not only of decay times but also DOP$_0$, and $\tau_\mathrm{r}$ values.
		
	\section{Discussion of results}\label{sec:results}
		We need to establish a correspondence between the theory and experimentally obtained DOP$_0$ and $\tau_i$ that were measured as a function of emission energy, not QDs dimensions, which are only roughly known, especially the length is uncertain. Additionally, as may be seen in Fig.~\ref{fig:f_dop_eta}(b), DOP$_0$ is virtually independent of the QD cross-section dimensions (for a given aspect ratio), hence the most straightforward interpretation, that QDs are all highly in-plane elongated and differ (from sample to sample) in cross-section size only immediately breaks down confronted with the relatively strong dispersion of experimentally obtained DOP$_0$.
		
		To express the theoretical results uniquely in terms of energy, we extrapolate the numerically obtained values of $E_\mathrm{X}=\left(E_V+E_H\right)/2$ and $\tau_i$ by fitting of smooth functions defined on the $H$-$L$ plane, and make an analytical mapping $\tau_i=f\lr{E_\mathrm{X}}$ by imposing a phenomenological relation between the size of the QD cross-section (defined by $H$) and $L$. It has been shown\cite{MarynskiJAP2013} that SEM images for unburied InAs/InAlGaAs/InP(001) nanostructures\cite{RudnoRudzinskiAPL2006} may be misleading, as changes in morphology after coverage with the barrier material can occur. To account for this, we use a simple relation of the form $L = a H^b$ as the QD geometry should scale mainly with the amount of deposited material. We do not aim here to establish a precise relation, but rather to find a qualitative trend that corresponds to agreement between experimental data and theory. In Fig.~\ref{fig:experiment}(b) and (c) we show the resulting theoretical curves for DOP$_0$ and $\tau_{V(H)}$, respectively, obtained for $a=\SI{22}{\nano\metre^{-1}}$, $b=2$. The resulting theoretical values of DOP$_0$ agree quantitatively with the experiment and very well reproduce the observed trend in the data. In the case of radiative lifetimes, the theory predicts well the qualitative character of the data, namely the dispersion of characteristic times of the two bright states. However, with generally shorter lifetimes, which may be partially attributed to a rather high uncertainty of the Kane energy parameter, $E_P$, which directly scales oscillator strengths. Additionally, in the diagonalization of the $\kp$ Hamiltonian a reduced $E_\mathrm{P}$ has to be used,\cite{VeprekPRB2007,BirnerBOOK2014} which results in reduced mixing of conduction and valence bands and hence smaller valence (conduction) band admixture to electron (hole) states. Such admixtures are dark, hence their reduction translates into overestimated oscillator strength values. These two effects may be, however, shown to affect both states equally in a simple manner (reduction of lifetimes by a common factor), which agrees with the fact that lifetimes for both states are underestimated, while DOP$_0$ describing their relation is well reproduced. Qualitatively analogous dispersion relations could be also obtained theoretically by assuming significant QDs shape fluctuations, which would be able to limit the confinement volume, as in smaller nanostructures. However, additional assumption on size of fluctuations scaling with QD emission energy would be needed. This may be addressed in terms of effective confinement length and as such is also captured by the discussion above.

		\begin{figure}[!tbp]
			\begin{center}
				\includegraphics[width=\columnwidth]{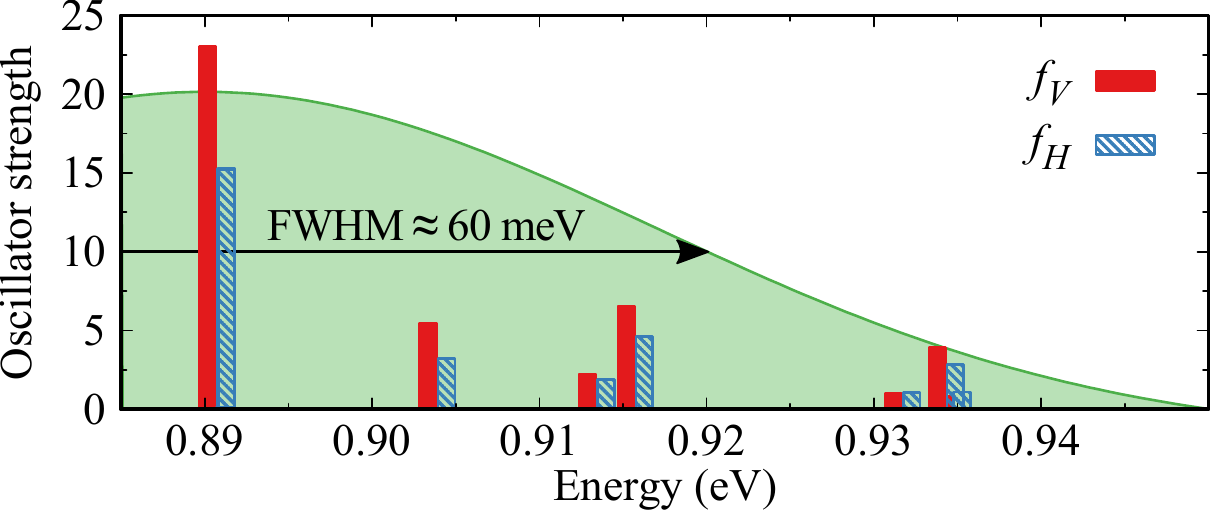}\upemhf
			\end{center}
			\caption{\label{fig:x_higher_f}(Color online) Calculated oscillator strengths $f_V$ (red full bars) and $f_H$ (blue striped bars) for a highly in-plane elongated ($\eta\approx5.8$) structure presented against a Gaussian distribution with FWHM typical for the width of PL bands from ensembles of investigated nanostructures (green shaded area).}
		\end{figure}
		A comment is needed on the measured DOP$_\mathrm{PL}$, as its values follow the identical dispersion trend but are generally lower from both experimentally obtained and theoretically calculated DOP$_0$. As it was discussed in Sec.~\ref{subsec:cw} and presented in Fig.~\ref{fig:cw_power}(b), in PL measurements the obtained DOP$_\mathrm{PL}$ may not fully reflect the intrinsic properties of exciton states, as the power-dependent occupation evolution plays an important role. In the performed PL measurements, identical excitation conditions have been maintained for all the samples (in each of the series with two different excitation powers), which however translates into different effective $\Gamma_P$ for each of them due to varying size (and hence areal density) of QDs. Indeed, using the above presumed relation between $L$ and $H$ (and consequently of both on $E_\mathrm{X}$) we may estimate the in-plane sizes to vary between \SI[product-units = single]{\sim11x70}{\nano\metre} and \SI[product-units = single]{\sim20x250}{\nano\metre}. This, assuming a maximally dense distribution of QDs, translates to planar densities of about \SIrange[range-units = single]{2e10}{1.3e11}{\centi\meter^{-2}} and approximately \SIrange{600}{4000}{} dots in the laser spot. We may therefore expect an almost order of magnitude mismatch of the effective power between the two ends of our total spectral window. This may explain a stronger decline in the measured DOP$_\mathrm{PL}$ as compared to DOP$_0$ as well as its generally lower values. Indeed, imposing the power-dependence of DOP$_\mathrm{PL}$ obtained for a mediocre simulated QD (which, after normalization in terms of DOP$_0$, represents a rather general character of this relationship) on theoretical values of DOP$_0$, allowed us to fit the theory also to DOP$_\mathrm{PL}$, which is presented in Fig.~\ref{fig:experiment}(b) via the dashed line. The theoretically anticipated power dependence of measured DOP$_\mr{PL}$ may be confirmed by comparison of the two data sets differing in excitation power (lighter and darker open symbols in Fig.~\ref{fig:experiment}(b)), with noticeably lower degree of polarization in the case of weaker excitation. Only S1 exhibits DOP$_\mr{PL}$ virtually immune to lowering power, which is in agreement with our initial assumption on existence of a WL-QD occupation transfer in this structure. As the polarization of luminescence depends on the capture rate of excitons, such additional channel that refills the exciton occupation strongly elevates the effective excitation power experienced by QDs.
		
		Additional slopes of various character present in DOP$_\mathrm{PL}$ for each of the samples may have multiple reasons. First, a possibly wide distribution of QDs length on a sample may lead to emission with varied degree of polarization correlated with the photon energy. Second, size and shape of QDs may have impact on the capture rate of excitons, which may be additionally modified by a possible exciton transfer between QDs within the ensemble. Finally, contribution of higher energy exciton states may be expected to be present within the main emission band (in contrast to well separated bands for ensembles of small QDs), mainly its high-energy tail, due to the dense energy ladder for elongated QDs. To confirm the latter, we present in Fig.~\ref{fig:x_higher_f} the calculated oscillator strengths ($f_V$ and $f_H$) for a highly elongated structure ($\eta\approx5.8$) against a Gaussian function with $\mr{FWHM}=\SI{60}{\milli\electronvolt}$, which is a typical shape of PL bands from investigated structures. One may notice that six additional reasonably bright states (the last two seem almost degenerate in this scale) fall into the PL band range, hence on its higher-energetic side one deals with mixed emission from exciton ground and excited states. Because of this, we restrict the reasoning presented here to the low-energy side of the emission band, where luminescence from the exciton ground state is expected to be dominant.
			
	\section{Conclusions}\label{sec:conclusions}
		We have investigated experimentally and theoretically the system of highly in-plane elongated quantum dots using as an exemplary system the InAs/Al$_{0.24}$Ga$_{0.23}$In$_{0.53}$As/InP(001) heterostructures grown by self-assembly in molecular beam epitaxy. The structures emit partially linearly polarized light in the spectral range of 1.25-$\SI{1.65}{\micro\meter}$, important in view of possible applications in telecom which may be tuned in the growth process by varying the amount of deposited material. Apart from the linear polarization of emission, structures exhibit a biexponential exciton recombination dynamics, with a significant dispersion for one of the decay times, which we have revealed both experimentally and theoretically with a good qualitative agreement. We show the underlying physics to be based on effects resulting from the structural asymmetry of the considered highly elongated QDs. First, the anisotropy of the confinement leads to a light-hole admixture to the nominally heavy-hole ground state in the valence band, which adds a contribution of opposite circular polarization to the standard emission of bright exciton. Second, the anisotropic electron-hole exchange interaction produces exciton eigenstates that couple to linearly polarized light in which the light-hole admixture causes a mismatch of oscillator strengths for the two axes of polarization. In the case of continuous wave excitation, this causes a strong power dependence of the degree of linear polarization, as predicted by our theoretical consideration and confirmed experimentally via polarization-resolved measurements under varied excitation power. In time-resolved spectroscopy in turn, the two components may be directly found, and a characteristic form of the biexponential decay is present, as under low excitation conditions each of the states emits the same amount of energy, and hence the ratios of amplitudes and decay times of the two components are connected. This allowed us to access experimentally the intrinsic polarization properties of the exciton ground state light-matter coupling, which occurred to be in quantitative agreement with the theory. The polarization of photoluminescence, that in the light of our findings does not fully reflect the characteristics of exciton ground state, indeed turned out to be generally lower than the \textit{intrinsic} degree of polarization, for which we successfully accounted theoretically based on the excitation power dependence of the emission kinetics. 
		
	\begin{acknowledgments}
		The work was supported by the Grant No. 2011/02/A/ST3/00152 from the Polish National Science Centre (Narodowe Centrum Nauki). K.~G. acknowledges support by the Grant No. 2014/13/B/ST3/04603 from the Polish National Science Centre (Narodowe Centrum Nauki). S.~H. acknowledges support from the State of Bavaria in Germany. Numerical calculations have been carried out using resources provided by Wroclaw Centre for Networking and Supercomputing (\url{http://wcss.pl}), Grant No. 203. M.~G. would like to thank Pawe{\l}~Machnikowski for helpful discussions.
	\end{acknowledgments}
		
%

\end{document}